**Amorphous Fe–Sn nanofilms for anomalous-Nernst heat-flux sensing**

*Kenji Tanabe*, Ko Mibu, Atsushi Tsukazaki and Kohei Fujiwara**

K. Tanabe 1
Toyota Technological Institute, Nagoya, Japan
E-mail: tanabe@toyota-ti.ac.jp

K. Mibu 2
Graduate School of Engineering, Nagoya Institute of Technology, Nagoya, Japan

A. Tsukazaki 3
Quantum-Phase Electronics Center (QPEC) and Department of Applied Physics, The University of Tokyo, Tokyo, Japan

A. Tsukazaki 3, K. Fujiwara 4
Institute for Materials Research, Tohoku University, Sendai, Japan

K. Fujiwara 4
Department of Chemistry, Rikkyo University, Tokyo, Japan
E-mail: kfujiwara@rikkyo.ac.jp

Funding: ARIM program of MEXT, Japan (No. JPMXP1225NI0401), Fuji Science and Technology Foundation, JST A-STEP (Grant No. JPMJTR25TA), JST FOREST (Grant No. JPMJFR223Y), JST CREST (Grant No. JPMJCR18T2), and the GIMRT Program of the Institute for Materials Research, Tohoku University (Proposal Nos. 202212-CRKEQ-0405 and 202402-CRKEQ-0046).

**Amorphous magnetic films are promising for anomalous-Nernst heat-flux sensing because their low thermal conductivity can enhance the temperature gradient generated by an applied heat flux. However, amorphization often degrades electronic transport and thermoelectric properties, making it challenging to obtain a large anomalous Nernst response in structurally disordered films. Here, we demonstrate nanometer-thick amorphous Fe–Sn films as high-sensitivity anomalous-Nernst heat-flux sensing materials. By systematically controlling composition and thickness, we find that amorphous Fe–Sn nanofilms combine a large anomalous Nernst response with low thermal conductivity, resulting in a heat-flux sensitivity of 0.37 μm $A^{-1}$. This value exceeds the sensitivities reported for both amorphous magnetic thin films and representative crystalline topological magnets. X-ray diffraction and Mössbauer spectroscopy show that the optimized films lack long-range crystallinity while retaining local Fe–Sn environments, suggesting that short-range atomic order contributes to the anomalous Nernst response in the amorphous matrix. The sensitivity is also reproduced on flexible polymer substrates, indicating compatibility with mechanically compliant device architectures. These results establish amorphous Fe–Sn nanofilms as a platform for anomalous-Nernst heat-flux sensing and provide a materials design route based on local-structure control and thermal-conductivity reduction.**

**1. Introduction**

Efficient thermal management is becoming increasingly important for next-generation energy, electronics, and sensing technologies, creating strong demand for high-sensitivity heat-flux sensors (HFSs). Among the available approaches, sensors based on the anomalous Nernst effect (ANE) [1-2] are particularly attractive because they are compatible with thin-film fabrication, flexible device geometries, and simple device structures. [3-10] These advantages make ANE-based HFSs promising for applications requiring lightweight, conformable, and spatially resolved thermal sensing.

The sensitivity of ANE-based HFSs is determined by the balance between transverse thermopower (V $K^{-1}$) and thermal conductivity (W $m^{-1}$ $K^{-1}$), and is expressed here in units of μm $A^{-1}$. From a materials-design perspective, the simultaneous realization of large transverse thermopower and low thermal conductivity is a prerequisite for achieving high sensitivity. In other words, materials with low thermal conductivity can sustain a larger temperature gradient under an applied heat flux, while large transverse thermopower enables the generation of a larger ANE voltage. Crystalline topological magnetic materials, including $Mn_3Sn$, [11] $Co_2MnGa$, [12-14] and $Fe_3Sn$, [15] have attracted considerable attention because Berry curvature in their electronic band structures gives rise to a large ANE. [11-26] However, these materials generally possess relatively high thermal conductivity due to their well-ordered crystal lattice, which limits their effectiveness for heat-flux sensing. By contrast, amorphous alloy nanofilms such as Gd-Co [8] intrinsically suppress phonon heat transport and can therefore achieve competitive sensitivity despite their more modest transverse thermopower.

Recent studies have further suggested that Berry-curvature-related transport phenomena can survive even in the absence of long-range crystalline order. [27-33] In particular, amorphous Fe–Sn alloy nanofilms have been shown to exhibit sizable transverse thermopower, implying that short-range structural motifs, such as kagome-related fragments, can sustain an intrinsic ANE response. [27] These findings raise the possibility of combining the low thermal conductivity of amorphous solids with the large transverse thermoelectric response generally believed to be specific to crystalline topological magnetic materials.

Here, we explore nanometer-thick amorphous $Fe_xSn_{1-x}$ films as a platform for high-sensitivity heat-flux sensing. By systematically investigating the composition dependence, we identify an optimal composition, $x = 0.74$, that achieves an exceptionally high sensitivity of +0.37 μm $A^{-1}$ at room temperature, exceeding previously reported values for ANE-based thin-film materials. We demonstrate that this high sensitivity originates from the synergistic effect of large transverse thermopower and suppressed thermal conductivity in the amorphous state. Mössbauer spectroscopy further reveals that approximately 30% of Fe atoms are associated with $Fe_3Sn$-like local environments, consistent with kagome-related short-range order. The high sensitivity is also preserved on flexible PEN substrates, highlighting the practical compatibility of this materials system with lightweight and conformable device platforms. These results establish amorphous Fe–Sn nanofilms as a promising functional material for high-performance flexible heat-flux sensing.

## 2. Results and discussion

### 2.1. Composition-Dependent Optimization of Heat-Flux Sensitivity

The sensitivity of amorphous $Fe_xSn_{1-x}$ films for ANE-based HFSs exhibits a clear maximum at $x = 0.74$. To evaluate this composition dependence under the heat-flux sensing configuration, we measured the ANE response in the geometry shown in **Figure 1(a)**, where the heat flux is applied perpendicular to the film plane. **Figure 1(b)** shows the dependence of the ANE electric field, $E_{ANE}$, on the external magnetic field, $\mu_0H$, and applied heat flux density, $j_Q^{Perp}$, for a $Fe_{0.74}Sn_{0.26}$ alloy film with a thickness $d$ = 35 nm. Due to the in-plane magnetic anisotropy of the $Fe_{0.74}Sn_{0.26}$ alloy film, the ANE electric field saturated in the high-magnetic-field region (> |5| mT), presenting a small hysteresis. The ANE electric field increased with increasing applied heat flux density. **Figure 1(c)** presents the relationship between the saturated ANE electric field taken at $\mu_0H = \pm 40$ mT in Figure 1(b), and the applied heat flux density. A clear linear relationship between $j_Q^{Perp}$ and $E_{ANE}$ is evident, and the slope, $E_{ANE}/ j_Q^{Perp}$, corresponds to the sensitivity of heat-flux sensing of the $Fe_{0.74}Sn_{0.26}$ alloy film. By applying linear fitting to the results for each $Fe_xSn_{1-x}$ film, the composition dependence of sensitivity was determined (**Figure 1(d)**). While the film thickness varied slightly with composition, it remained within the range of $d$ = 33–37 nm. The sensitivity tended to increase with the Fe content reaching a maximum of +0.37 μm $A^{-1}$ at $x = 0.74$ before turning to decrease. Interestingly, the composition of $x = 0.74$ closely corresponds to the crystalline $Fe_3Sn$ composition with a kagome lattice, suggesting the positive contribution of short-range kagome-lattice order in these amorphous films.

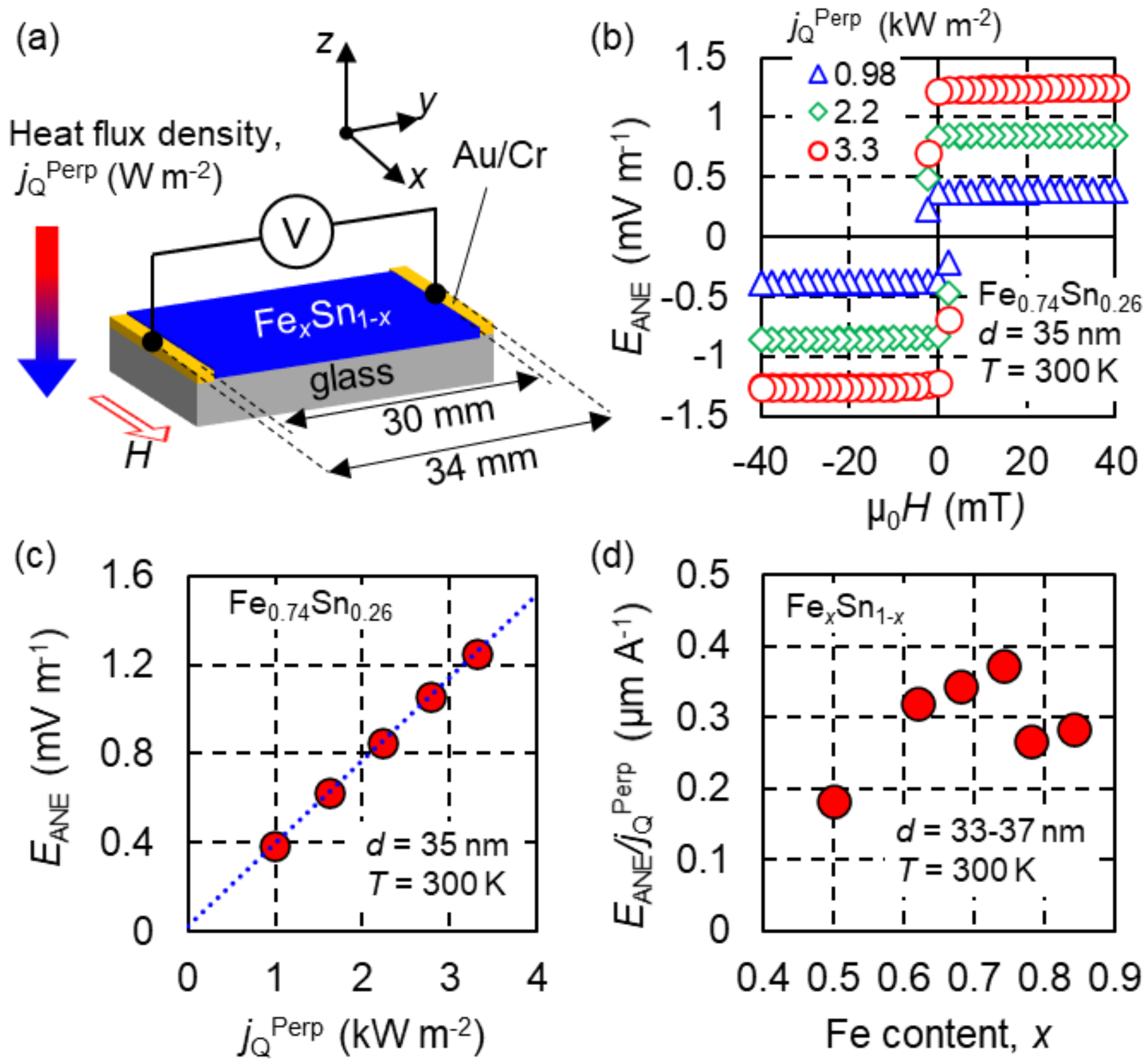

**Figure 1 | Composition-dependent optimization of sensitivity of heat-flux sensing in amorphous $Fe_xSn_{1-x}$ films.**

(a) Schematic illustration of the measurement geometry. (b) ANE electric field, $E_{ANE}$, as a function of external magnetic field for a 35-nm-thick $Fe_{0.74}Sn_{0.26}$ film at room temperature under different heat flux densities, $j_Q^{Perp}$. (c) Saturated $E_{ANE}$ plotted as a function of $j_Q^{Perp}$, where the value at $\mu_0 H = \pm 40$ mT was extracted from (b). The dotted line is a linear fit, and its slope gives the sensitivity. (d) Sensitivity, $E_{ANE}/j_Q^{Perp}$, as a function of Fe content, $x$.

To benchmark the performance of amorphous Fe–Sn thin films against representative ANE-based heat-flux-sensing materials, we compared the sensitivity obtained here with previously reported values for thin-film systems (**Figure 2**). Amorphous $Fe_{0.74}Sn_{0.26}$ exhibits the highest sensitivity among the compared thin-film materials, exceeding the previously reported values for epitaxial $Co_2MnGa$, [4] $Fe_3Sn$/Pt, [7] and amorphous $Gd_{0.24}Co_{0.76}$, [8] and thereby demonstrating a substantial improvement in heat-flux-sensing performance.

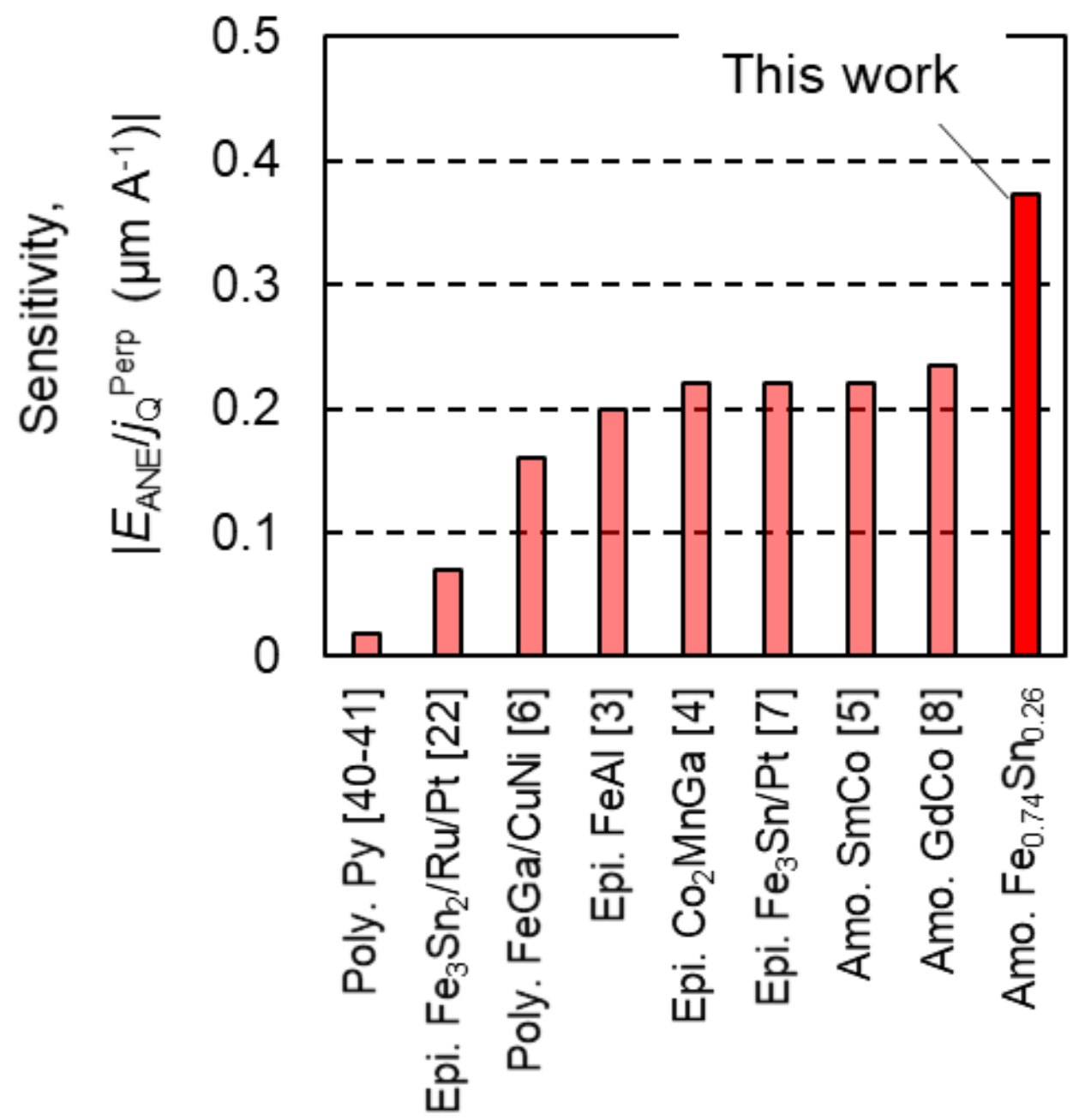

**Figure 2 | Benchmarking the sensitivity of representative thin-film ANE-based heat-flux sensing materials.**

Sensitivities collected from the literature and from the present work. The values were obtained or estimated from reported ANE voltage, transverse thermopower, thermal conductivity, and device geometries. Data for polycrystalline Py [40,41] were calculated from the reported $S_{ANE}$ and $\kappa$. Data for polycrystalline FeGa/CuNi, [6] epitaxial FeAl, [3] epitaxial $Co_2MnGa$, [4] and amorphous SmCo [5] were estimated from the reported $V_{ANE}/j_Q^{Perp}$ and device structures. Amorphous Fe–Sn exhibits the highest sensitivity among the compared thin-film systems.

**2.2. Origin of the High Sensitivity**

To clarify the materials origin of the high sensitivity observed in the device measurements, we next investigated the transport properties of amorphous Fe–Sn alloy films in a conventional lateral measurement geometry. Thin films identical to those used in Figure 1 were measured. As shown in **Figure 3(a)**, the transverse thermopower was measured along *y* direction under conditions where the temperature gradient was applied along the in-plane *x* direction and the magnetic field was applied along the out-of-plane *z* direction. **Figure 3(b)** shows the composition dependence of the transverse thermopower, $S_{ANE}$. The transverse thermopower exhibited a monotonic increase with increasing Fe content, consistent with prior research results. [27] **Figure 3(c)** illustrates the composition dependence of longitudinal resistivity, $\rho_{xx}$. The resistivity tends to decrease with increasing Fe content.

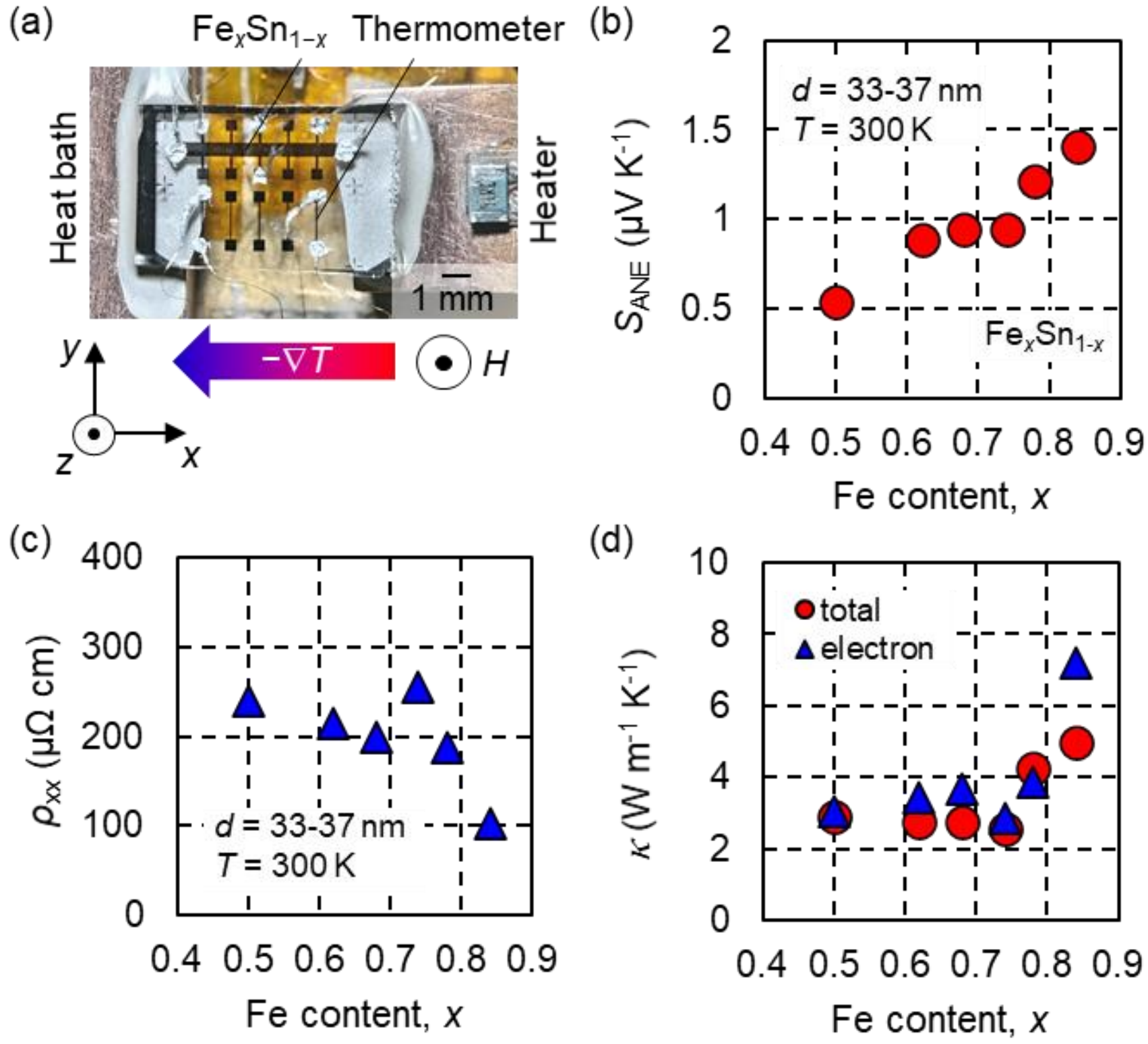

**Figure 3 | Composition dependence of transport properties in amorphous Fe–Sn films.** (a) Photograph of the measurement setup and measurement geometry. (b) Transverse thermopower, $S_{ANE}$, as a function of Fe content at room temperature for the films with thicknesses of 33–37 nm. (c) Longitudinal resistivity, $\rho_{xx}$, as a function of Fe content. (d) Total and electronic thermal conductivities as functions of Fe content. The total thermal conductivity was evaluated from $\kappa = S_{xy}/(E_{ANE}/j_Q^{Perp})$, and the electronic thermal conductivity was estimated using the Wiedemann–Franz law, $\kappa_{electron} = L\sigma_{xx}T$. For the transport quantities related to the ANE response, the values were evaluated from data averaged over the magnetic-field region of $\mu_0 H = 2.5$–$3.0$ T.

Using the analysis method proposed by Odagiri et al., the thermal conductivity was evaluated as follows. [8] Sensitivity can be expressed as the ratio of transverse thermopower to total thermal conductivity that directly links to the degree of heat-flux density. Conversely, thermal conductivity can be expressed as the ratio of transverse thermopower to sensitivity. Here, the transverse thermopower corresponds to $S_{yz}$, measured with an out-of-plane temperature gradient and an in-plane magnetic field. Assuming the amorphous nature of the Fe-Sn alloy eliminates this transport anisotropy, we take $S_{yz} = S_{xy}$. Under this assumption, the thermal conductivity can be evaluated as $S_{xy}/(E_{ANE}/j_Q^{Perp})$. The red circles in **Figure 3(d)** represent this result, showing that thermal conductivity, $\kappa$, was suppressed to approximately $\kappa$ = 3 W m$^{-1}$ K$^{-1}$. This value is nearly one order of magnitude smaller than the room-temperature thermal conductivity reported for representative crystalline ANE materials such as $Co_2MnGa$ (~20 W m$^{-1}$ K$^{-1}$), highlighting the advantage of the amorphous state for heat-flux sensing. Below $x$ = 0.74, thermal conductivity remained relatively constant, indicating that the increased sensitivity in this region (Figure 1(d)) originated from the enhanced transverse thermopower. Above $x$ = 0.74, the sensitivity decrease (Figure 1(d)) was attributed mainly to an increase in thermal conductivity. These results indicate that the superior sensitivity of amorphous Fe–Sn in Figure 2 arises not only from its large ANE response but also from its exceptionally low thermal conductivity.

Using the Wiedemann-Franz law, the contribution of conduction electrons to thermal conductivity was evaluated. According to this law, the thermal conductivity of conduction electrons is given by $\kappa = L\sigma_{zz}T$, where $L$ is the Lorenz number and $\sigma_{zz}$ is the electrical conductivity along the $z$-axis. Assuming the isotropic transport again, i.e., $\sigma_{zz} = \sigma_{xx}$, and using resistivity data from Figure 3(c) and Hall resistivity data (not shown), the electrical conductivity, $\sigma_{xx}$, was calculated as $\sigma_{xx} = \rho_{xx}/\left(\rho_{xx}{}^2 + \rho_{xy}{}^2\right)$, with $\rho_{xy}$ being the Hall resistivity. Here, $\rho_{xy}$ was taken from the magnetization-saturated high-field region ($\mu_0 H$ = 2.5–3.0 T) for consistency with the ANE-related quantities. The blue triangles in Figure 3(d) represent the result, which overall aligns well with the total thermal conductivity, although some scatter is observed. Consequently, it is concluded that the phonon contribution of thermal conductivity is suppressed by the amorphous nature of the material. Although the evaluated electronic thermal conductivity slightly exceeds the total thermal conductivity at some compositions, this small apparent inversion is attributed to the uncertainty associated with the indirect evaluation procedures and does not affect the overall conclusion that the phonon contribution is strongly suppressed. The decrease in sensitivity above $x$ = 0.74 was attributed to the increased thermal conductivity of conduction electrons.

### 2.3. Thickness Optimization

The sensitivity of $Fe_{0.78}Sn_{0.22}$ films is also strongly dependent on film thickness, with an optimum around $d$ = 33 nm. As shown in **Figure 4(a)**, sensitivity increased with film thickness up to 33 nm but decreased beyond this value. The Fe content was fixed at $x$ = 0.78 for this characterization. Using the same analytical methods as in Figure 3, the transverse thermopower (**Figure 4(b)**), longitudinal resistivity (**Figure 4(c)**), and thermal conductivity were evaluated. **Figure 4(d)** shows the thickness dependence of thermal conductivity. The phonon contribution to thermal conductivity was calculated by subtracting the conduction electron contribution from the total thermal conductivity. For thicknesses greater than 33 nm, thermal conductivity increased, with a notable rise in the phonon contribution. This result indicates that the low-thermal-conductivity advantage of the amorphous state is gradually weakened in thicker films, suggesting an accompanying structural change.

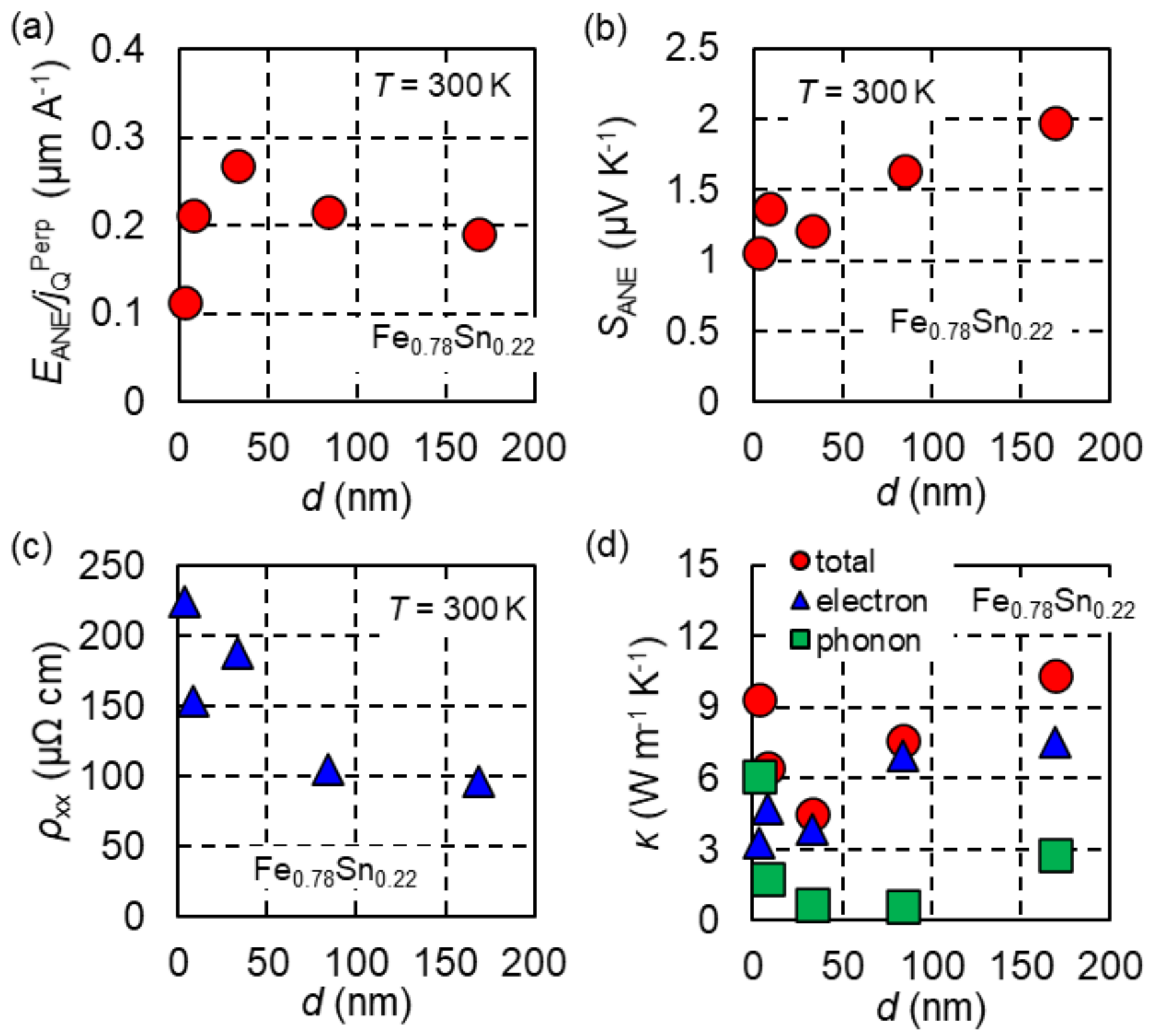

**Figure 4 | Thickness-dependent optimization of transport properties in amorphous Fe–Sn films.**

(a-c) Sensitivity (a), transverse thermopower (b), and longitudinal resistivity (c) as a function of thickness. The Fe content is fixed at $x = 0.78$. The sensitivity values in (a) were evaluated from the ANE electric field measured at $\mu_0 H = 40$ mT, whereas the transverse thermopower values in (b) were obtained from data averaged over the magnetic-field range of $\mu_0 H = 2.5$–3.0 T. (d) Total (red circles), electronic (blue triangles), and phonon (green squares) thermal conductivities as functions of film thickness. The total thermal conductivity was evaluated from the sensitivity and transverse thermopower. The electronic thermal conductivity was calculated using the Wiedemann-Franz law. The phonon thermal conductivity is the difference between the total and electronic thermal conductivities.

**2.4. Structural Origin of the Thickness Dependence**

To examine the structural origin of the thickness-dependent change, X-ray diffraction (XRD) measurements were conducted (**Figure 5(a)**). No crystalline peaks were observed below 33 nm, but a weak peak near diffraction angle $2\theta = 43°$ appeared for films thicker than 85 nm, becoming more pronounced at 170 nm. This weak peak can be indexed to (110) of a considerably disordered bcc-Fe(Sn) solid solution. It might exist as nanocrystals in the amorphous matrix, which could increase both phonon and conduction electron thermal conductivity, leading to reduced sensitivity (Figure 4(a)). This phenomenon may be attributed to a slight increase in substrate temperature during extended deposition times, promoting crystallization. From the viewpoint of heat-flux sensing, this behavior emphasizes that maintaining the amorphous state is a key materials requirement. Although crystalline and epitaxial magnetic films are often regarded as favorable platforms for large ANE, our results show that the low-thermal-conductivity advantage of the amorphous state is essential for achieving high sensitivity in Fe–Sn nanofilms.

Additionally, the thermal conductivity showed an increasing trend below $d = 8.6$ nm. In contrast to the thicker films, no crystalline peaks were detected by XRD in this region, suggesting a different origin from the behavior observed above 33 nm. At present, the origin of this increase remains unclear. It may reflect an additional contribution that becomes important in the ultrathin region; for example, the assumption of isotropy used in the present analysis may no longer be fully valid at very small thicknesses.

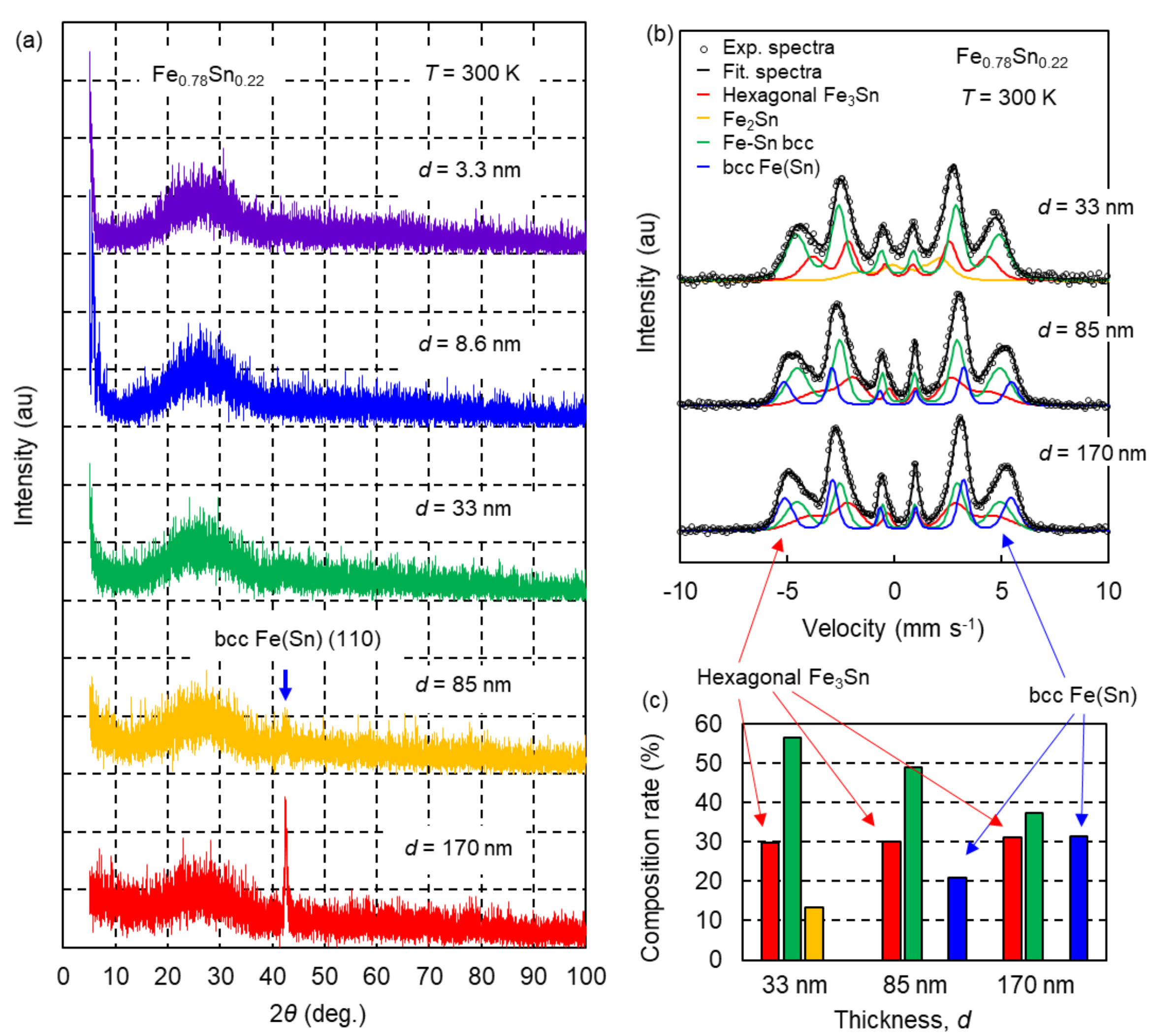

**Figure 5 | Structural characterization of $Fe_{0.78}Sn_{0.22}$ films.**
(a) X-ray diffraction patterns of $Fe_{0.78}Sn_{0.22}$ films with thicknesses of 3.3, 8.6, 33, 85, and 170 nm at room temperature. (b) $^{57}$Fe Mössbauer spectra of $Fe_{0.78}Sn_{0.22}$ films at room temperature. Open circles represent the experimental data and black curves represent the total fits. The fitted components are assigned as described in Note S2. (c) Thickness dependence of the fitted fractions of the local structural components extracted from the Mössbauer analysis.

### 2.5. Local Structural Analysis by Mössbauer Spectroscopy

To gain microscopic insight into the local Fe environments, we performed $^{57}$Fe Mössbauer spectroscopy on $Fe_{0.78}Sn_{0.22}$ films with thicknesses of $d$ = 33, 85, and 170 nm (see Note S1). The spectra exhibit relatively well-resolved magnetic splitting, which is unusual for a fully random magnetic amorphous alloy (**Figure 5(b)**). We analyzed the spectra using components assigned on the basis of reported Mössbauer spectra of related Fe–Sn phases (see Note S2). [34-

[38] The fitting results reveal that all films contain an $Fe_3Sn$-like local component with an approximately constant fraction of ~30%, while the bcc Fe(Sn)-like contribution increases with increasing thickness (**Figure 5(c)** and Table S1). The presence of the $Fe_3Sn$-like local component is consistent with previous extended X-ray absorption fine structure (EXAFS) results showing that amorphous Fe–Sn films retain kagome-related short-range order at the nearest-neighbor length scale. [27] By contrast, the thickness-dependent increase in the bcc Fe(Sn)-like contribution is consistent with the structural evolution inferred from the XRD results and supports the development of partial crystallization in thicker films. The persistence of the $Fe_3Sn$-like local component suggests that kagome-related short-range order survives in the amorphous matrix and may contribute to the large ANE observed in these films.

**2.6. Flexible Heat-Flux Sensing on PEN**

Finally, we present the results of fabricating Fe-Sn nanofilms on a PEN substrate for potential application in HFSs. Most commercially available HFSs are fabricated as devices on thin plastic substrates. Since amorphous Fe-Sn alloy films require neither high-temperature deposition nor annealing to show the large ANE, they are expected to exhibit similarly high sensitivity on plastic substrates. To evaluate this, Fe-Sn alloy films ($d$ = 40 nm) were deposited on a PEN substrate with a thickness of 100 μm, and their sensitivity was measured as shown in **Figure 6(a-c)**. The sensitivity was found to be consistent with that of Fe-Sn nanofilms deposited on glass substrates, demonstrating that the films function equivalently on plastic substrates (Figure 6(c)). This result indicates that amorphous Fe-Sn alloy films not only possess exceptionally high sensitivity but also exhibit excellent properties from an application perspective, as they can be deposited on plastic substrates, offering significant potential for practical HFS applications.

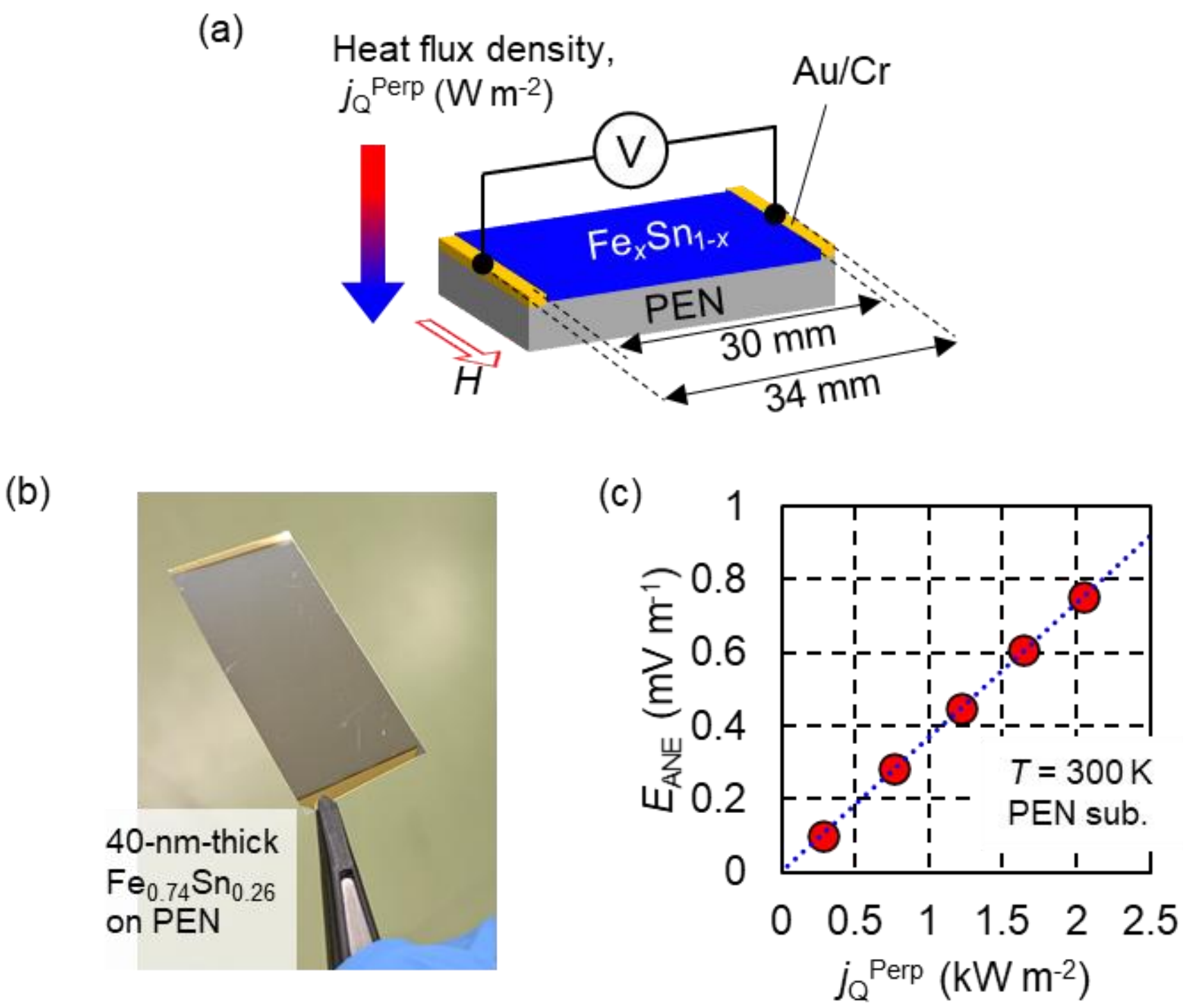

**Figure 6 | Sensitivity of amorphous Fe–Sn films on a flexible PEN substrate.**
(a) Schematic illustration of the measurement geometry. (b) Photograph of the device on a PEN substrate. (c) ANE electric field plotted as a function of $j_Q^{Perp}$ for a 40-nm-thick $Fe_{0.74}Sn_{0.26}$ film on PEN at room temperature. The dotted line is a linear fit, and its slope gives the sensitivity, consistent with the value obtained on glass substrates (Figure 1(d)).

## 3. Conclusions

In summary, we investigated amorphous Fe–Sn alloy nanofilms as a new materials platform for anomalous-Nernst-effect-based heat-flux sensing. We identified an optimal composition of $Fe_{0.74}Sn_{0.26}$ and a film thickness of ~35 nm, at which the sensitivity reached +0.37 μm $A^{-1}$, exceeding previously reported values for ANE-based thin-film materials. This exceptionally high sensitivity originates from the combination of large transverse thermopower and low thermal conductivity in the amorphous nanofilms.

Composition- and thickness-dependent analyses revealed that the sensitivity is maximized when the transverse thermopower is enhanced while the thermal conductivity remains suppressed. In thicker films, the sensitivity decreases because partial crystallization increases the thermal conductivity, particularly its phonon contribution. Mössbauer spectroscopy further showed that the amorphous Fe–Sn nanofilms contain $Fe_3Sn$-like local environments with an approximately constant fraction of ~30%, providing microscopic support for kagome-related short-range order in the amorphous matrix.

In addition to their high sensitivity, amorphous Fe–Sn nanofilms retain their performance on flexible PEN substrates and can be deposited at room temperature without post-annealing, highlighting their practical compatibility with lightweight and conformable device platforms. These results establish amorphous Fe–Sn nanofilms as a promising functional material for high-performance flexible heat-flux sensing and further motivate the concept of amorphous topological magnets.

**4. Experimental Section/Methods**

*Fabrication*:

We fabricated Si–O(15 nm)/$Fe_xSn_{1-x}$($d$) nanofilms on glass substrates (Matsunami Glass Ind., Ltd. S1126) at room temperature using a radio-frequency magnetron sputtering system. Fe-Sn alloy films were deposited by co-sputtering. [39] The samples on glass substrates used for the sensitivity measurement were 34 × 17 $mm^2$ in planar dimensions, and Au/Cr strip-type electrodes (2 mm in length, 17 mm in width) were electron-beam-deposited on the edges of the substrate before depositing the Fe-Sn film (Figure 1(a)). In order to measure the transport properties, the nanofilm samples were patterned to a Hall-bar shape using photolithography and Ar ion milling (Figure 3(a)).

*Structural Characterization*:

The composition was determined by energy dispersive X-ray analysis. XRD measurements using Cu $K\alpha_1$ radiation were performed for the structural characterizations. The thickness of the amorphous Fe-Sn nanofilms ranged from 3.3 to 170 nm, which was calculated from the fitting analysis of X-ray reflectivity data.

*Transport measurements*:

The sensitivity of ANE-HFSs was measured by applying the heat flux in the out-of-plane direction and the magnetic field in the in-plane direction. This method is often termed the heat-flux method. [40] Details of the measurement setup are described in previous papers. [8] The evaluation procedure for the applied heat flux density is summarized in Note S4. To ensure the measurement accuracy, the setup was calibrated using Py films. [41-42] In contrast, the transverse thermopower of the ANE was measured by applying heat flux in the in-plane direction and the magnetic field in the out-of-plane direction. Using a Hall-bar sample, longitudinal resistivity, Hall resistivity, and longitudinal thermopower were measured. All measurements were conducted at room temperature.

*Mössbauer measurements*:

$^{57}$Fe Mössbauer spectra were acquired at room temperature using a conventional constant-acceleration spectrometer in standard scattering geometry with conversion electron detection using a $^{57}$Co source. Detailed experimental conditions, fitting procedures, and the assignment of the fitted components are described in Note S1 and Note S2.

**Acknowledgements**

The authors thank Ms. Onoue for assistance with the Mössbauer measurements in a framework of ARIM program of MEXT, Japan (No. JPMXP1225NI0401). We thank Dr. J. Shiogai for helpful discussions. This work was supported by multiple funding sources, including the Fuji Science and Technology Foundation, JST A-STEP (Grant No. JPMJTR25TA), JST FOREST (Grant No. JPMJFR223Y), and JST CREST (Grant No. JPMJCR18T2). Part of this work was conducted under the GIMRT Program of the Institute for Materials Research, Tohoku University (Proposal Nos. 202212-CRKEQ-0405 and 202402-CRKEQ-0046).

**Data Availability Statement**

The data that support the findings of this study are available from the corresponding author upon reasonable request.

**Supporting Information**

Supporting Information is available from the Wiley Online Library or from the author.

Supporting Information

**Amorphous Fe–Sn nanofilms for anomalous-Nernst heat-flux sensing**

*Kenji Tanabe*, Ko Mibu, Atsushi Tsukazaki and Kohei Fujiwara**

## Note S1. $^{57}$Fe Mössbauer measurements of amorphous Fe–Sn films

To obtain microscopic information on the local Fe environments in amorphous Fe–Sn films, we performed room-temperature $^{57}$Fe Mössbauer spectroscopy on three amorphous $Fe_{0.78}Sn_{0.22}$ films with thicknesses of 33, 85, and 170 nm. As a crystalline reference, a 25-nm-thick $Fe_3Sn$ film grown on Pt/$Al_2O_3$(0001) was also measured. The amorphous films were deposited on glass substrates with an Si–O capping layer, whereas the reference film had the structure Si–O(15 nm)/crystalline $Fe_3Sn$/Pt(4 nm)/$Al_2O_3$(0001).

The measurements were carried out at room temperature using conversion-electron Mössbauer spectroscopy (CEMS) in a scattering geometry. A He-based counter gas was used for conversion-electron detection. The Doppler velocity scale was calibrated using an α-Fe foil, and the centroid of pure Fe was defined as 0 mm $s^{-1}$. Measurement durations were 26 days for the 33-nm amorphous film, 3 days for the 85- and 170-nm amorphous films, and 16 days for the crystalline $Fe_3Sn$ reference film.

All spectra exhibit magnetic hyperfine splitting at room temperature, indicating that magnetically ordered phases dominate in both the amorphous and crystalline samples. In addition, the peak-intensity ratios are close to 3:4:1:1:4:3, suggesting that the magnetic moments lie predominantly in the film plane. Compared with spectra typical of fully random magnetic amorphous materials, the present spectra show relatively narrow line broadening and well-defined sextet-like features, implying that specific local structural motifs survive even in the amorphous Fe–Sn films.

## Note S2. Fitting procedure for the Mössbauer spectra

Because all spectra show magnetic splitting with relatively well-resolved peaks, we analyzed them using the minimum number of magnetically split subspectra required to reproduce the data. Each absorption peak was modeled by a Voigt profile, that is, a Lorentzian line shape convoluted with a Gaussian distribution of hyperfine field. Since several magnetically split components overlap continuously in each spectrum, the uniqueness of the fitting is inherently limited. To improve stability, the Lorentzian linewidth and the sextet intensity ratio 3:x:1:1:x:3 were constrained to be common among the subspectra within each spectrum. For clarity, Fe(1)–Fe(5) denote fitted subspectral components and do not necessarily correspond to single crystallographic phases.

For the 85- and 170-nm amorphous $Fe_{0.78}Sn_{0.22}$ films, the spectra were fitted using three components, denoted Fe(1), Fe(2), and Fe(3). The Fe(1) component, with a hyperfine field of about 25 T, was assigned to an Fe site with a local environment similar to that of hexagonal $Fe_3Sn$ [1-2]. Two additional components with larger hyperfine fields, Fe(2) near 29 T and Fe(3) near 33 T, were introduced to reproduce Fe-rich cubic or bcc-like local environments with different local Sn coordinations [1]. For Fe(2) and Fe(3), the quadrupole shift was fixed at zero. The spectrum of the 33-nm amorphous film differs from those of the thicker amorphous films. In this sample, the intensity near −6.0 to −4.3 mm $s^{-1}$ and +4.9 to +6.0 mm $s^{-1}$ is reduced, while the central part of the spectrum is enhanced. The best fit was therefore obtained using Fe(1), Fe(2), and an additional low-field component Fe(4) with a hyperfine field around 12 T. This Fe(4) component was interpreted as a local environment reminiscent of FeSn- or $FeSn_2$-like phases [3-4]. In contrast, the Fe(3) component associated with the most Fe-rich local environment was not required for the 33-nm film.

For the crystalline $Fe_3Sn$ reference film, the spectrum was fitted using Fe(1), Fe(3), and a broad Fe(5) component around 20 T. The Fe(5) component was regarded as consistent with $Fe_5Sn_3$- or $Fe_3Sn_2$-like contributions [5]. This analysis indicates that the reference film is not a single-phase hexagonal $Fe_3Sn$ sample, but contains additional coexisting phases.

## Note S3. Fitting results and implication for the amorphous Fe–Sn films

The fitted area fractions reveal that the Fe(1) component, assigned to an $Fe_3Sn$-like local structure, accounts for 29.9%, 30.1%, and 31.3% of the total spectral area in the 33-, 85-, and 170-nm amorphous $Fe_{0.78}Sn_{0.22}$ films, respectively. Thus, approximately 30% of Fe atoms are associated with $Fe_3Sn$-like local environments, and this fraction is nearly independent of thickness over the investigated range.

At the same time, the remaining spectral weight is dominated by non-$Fe_3Sn$-like components. In the 85- and 170-nm films, Fe(2) and Fe(3), which represent Fe-rich cubic or bcc-like local environments, together occupy most of the spectral area. In the 33-nm film, Fe(2) remains dominant and Fe(4), a low-field component, also appears. These results indicate that the amorphous Fe–Sn films are microscopically heterogeneous and cannot be described as a single $Fe_3Sn$-derived local structure.

The Mössbauer results nevertheless provide direct microscopic support for the presence of $Fe_3Sn$-like short-range order embedded in the amorphous matrix. This interpretation is consistent with the large anomalous Nernst response observed in the amorphous Fe–Sn films and supports the discussion in the main text that kagome-related short-range order can survive in an amorphous system. In addition, the increasing fraction of bcc-like components in thicker films is consistent with the emergence of a bcc-Fe(Sn)-related diffraction peak in X-ray diffraction, as discussed in the main text.

## Note S4. Calibration of the heat-flux sensing measurement

To evaluate the heat-flux density applied to the sample, we used a commercial heat-flux sensor placed beneath the substrate, as schematically shown in Figure S3. In this geometry, the measured heat flux corresponds to the heat conducted through the substrate and detected by the commercial sensor. Because part of the supplied heat can be lost through thermal radiation and natural convection, the detected heat flux may be smaller than the heat flux passing through the sample.

To assess this effect, we performed a calibration measurement using a Py reference film prepared on the same glass substrate and measured in the same configuration as the Fe–Sn films, with only the magnetic thin film replaced by Py. Since the substrate and thermal stack were identical, the heat loss associated with radiation and convection is expected to be nearly the same as that in the Fe–Sn measurements, allowing the reference measurement to calibrate the effective heat flux passing through the substrate.

The measured result for the Py film is shown in Figure S3(b). From the measured slope, the sensitivity was evaluated to be +0.021 μm/A, which is approximately 15% larger than the value expected from previously reported Py data (+0.018 μm/A [6-7]). Because the sensitivity is defined as ($E_{ANE}/j_Q^{Perp}$), this larger measured value indicates that the heat-flux density detected by the commercial sensor is underestimated relative to the actual heat flux passing through the sample. This underestimation is most likely caused by heat loss through radiation and convection. Nevertheless, because the calibration and Fe–Sn measurements were performed using the same substrate and thermal configuration, this systematic reduction is expected to be largely compensated in the present evaluation.

To further validate the reference Py film independently of the heat-flux calibration, we measured the transverse thermopower of the same Py film in the in-plane temperature-gradient geometry shown in Figure S3(c). The obtained value (+0.54 μV/K) was consistent with previously reported Py data [6], indicating that the Py film itself was properly prepared and that the discrepancy in the heat-flux calibration mainly originates from the estimation of heat-flux density.

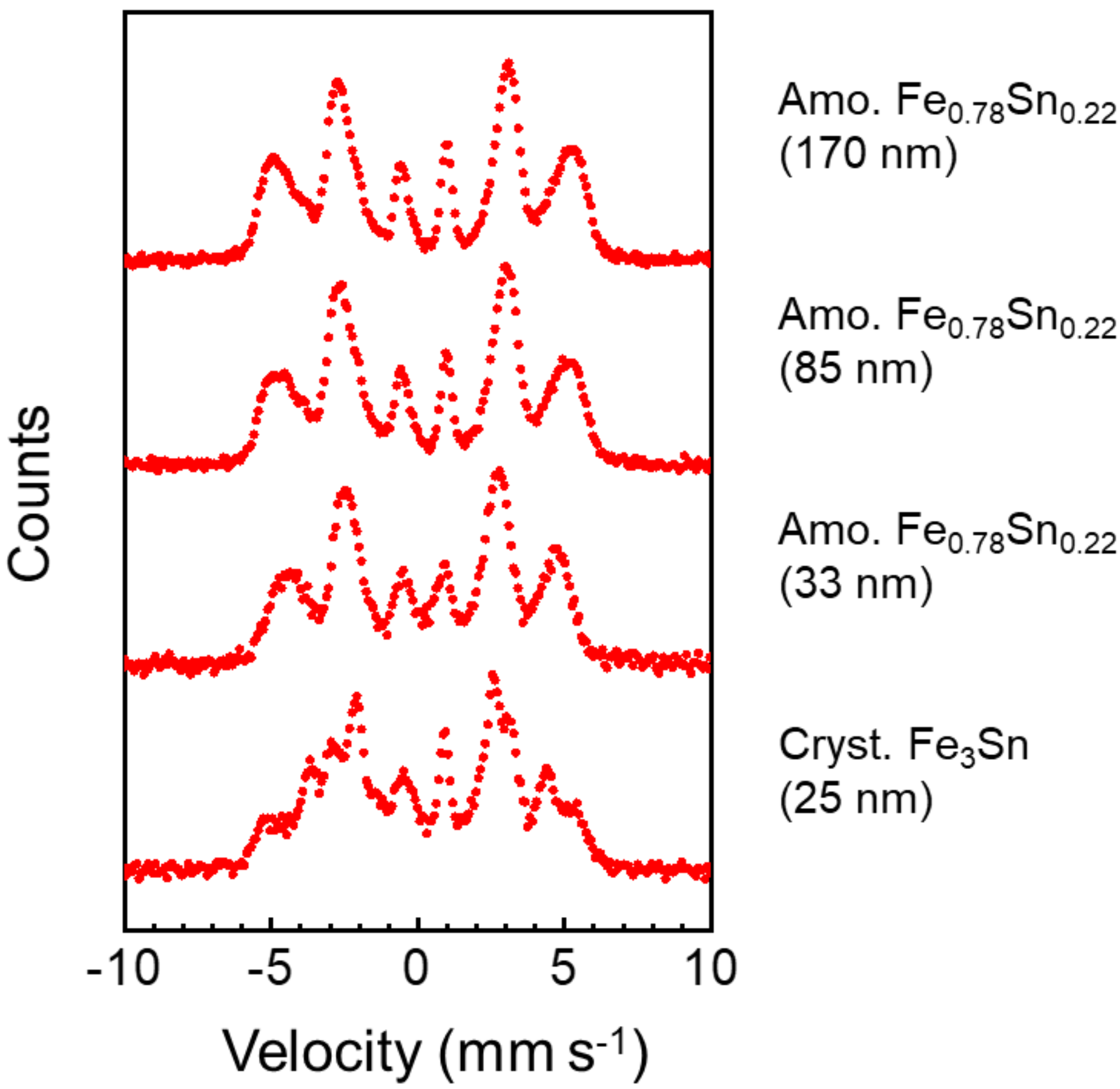

**Figure S1 | Room-temperature $^{57}Fe$ conversion-electron Mössbauer spectra of amorphous and crystalline Fe–Sn films.**
Measured spectra for amorphous $Fe_{0.78}Sn_{0.22}$ films with thicknesses of 33, 85, and 170 nm, together with a crystalline $Fe_3Sn$ reference film with a thickness of 25 nm. All spectra show magnetic hyperfine splitting.

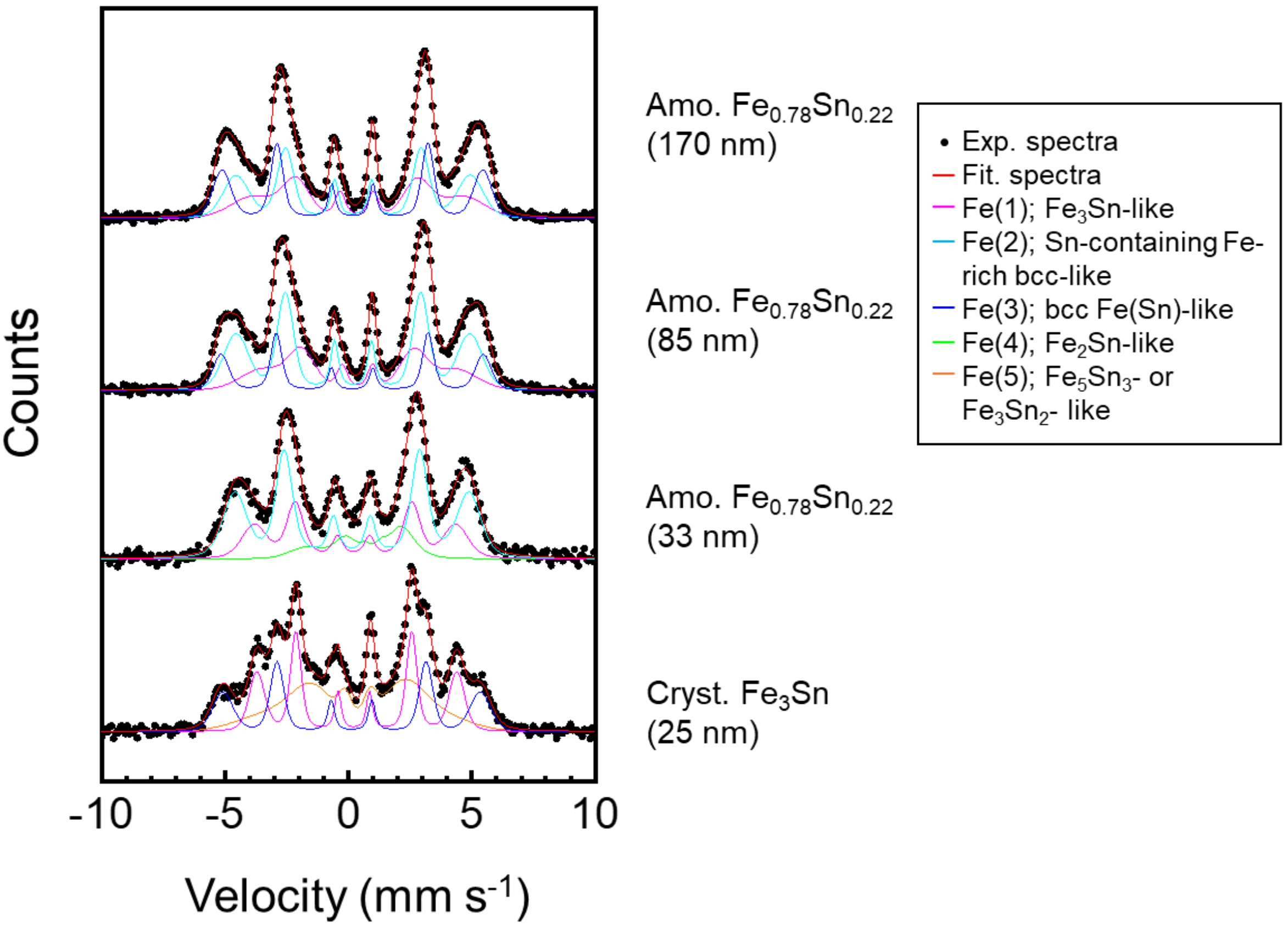

**Figure S2 | Best-fit decomposition of the room-temperature Mössbauer spectra.** Black circles represent the experimental data, red curves represent the total fits, and colored curves show the individual magnetically split subspectra used in the fitting. The amorphous films were fitted with three components each, while the crystalline $Fe_3Sn$ reference film was fitted with three different components as described in Note S2.

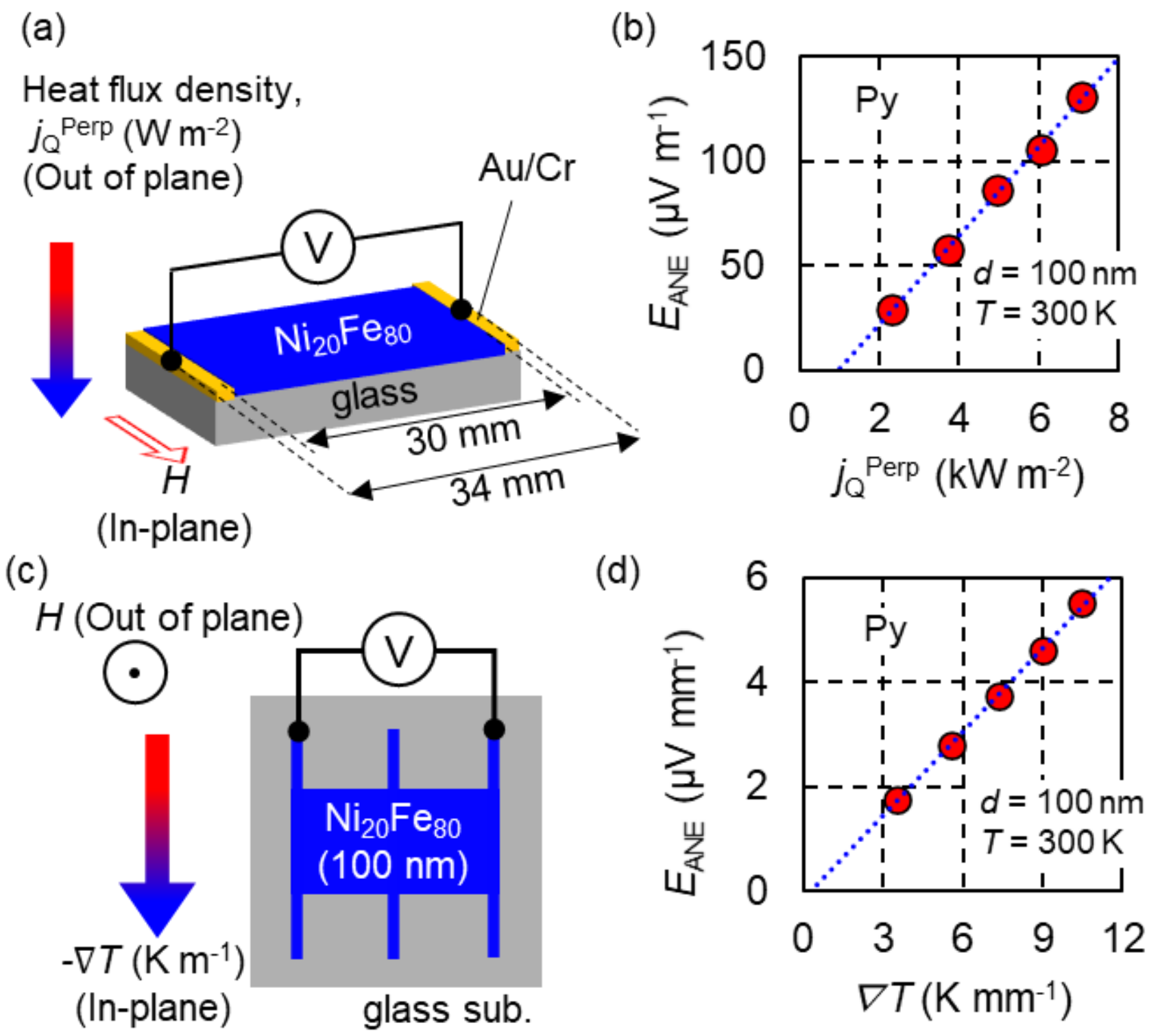

**Figure S3 | Calibration of the heat-flux sensing setup and validation of the Py reference film.**
(a) Schematic illustration of the heat-flux sensing geometry used for calibration of the setup. (b) ANE electric field ($E_{ANE}$) as a function of perpendicular heat-flux density ($j_Q^{Perp}$) for a 100-nm-thick Py film at 300 K. The dotted line is a linear fit. (c) Schematic illustration of the in-plane temperature-gradient geometry used to independently evaluate the transverse thermopower of the Py reference film. (d) ANE electric field ($E_{ANE}$) as a function of in-plane temperature gradient ($\nabla T$) for the same Py film at 300 K. The linear response and the obtained value are consistent with previously reported Py data, supporting the validity of the reference film and the measurement procedure.

* Fixed during fitting.
† Common fitting parameters.

Amorphous $Fe_{0.78}Sn_{0.22}$, 170 nm

| Component | Assignment | δ (mm s$^{-1}$) | 2ε (mm s$^{-1}$) | $B_{hf}$ (T) | Γ (mm s$^{-1}$) | Δ$B$ (T) | ($x$) in 3:x:1:1:x:3 |
|---|---|---|---|---|---|---|---|
| Fe(1) | $Fe_3Sn$-like local environment | 0.37 | 0.06 | 26.0 | †0.30 | 5.2 | †3.33 |
| Fe(2) | Sn-containing Fe-rich bcc-like local environment | 0.20 | *0.00 | 29.3 | †0.30 | 2.8 | †3.33 |
| Fe(3) | bcc Fe(Sn)-like local environment | 0.16 | *0.00 | 32.7 | †0.30 | 1.9 | †3.33 |

($\chi^2$ = 1.67)

Amorphous $Fe_{0.78}Sn_{0.22}$, 85 nm

| Component | Assignment | δ (mm s$^{-1}$) | 2ε (mm s$^{-1}$) | $B_{hf}$ (T) | Γ (mm s$^{-1}$) | Δ$B$ (T) | ($x$) in 3:x:1:1:x:3 |
|---|---|---|---|---|---|---|---|
| Fe(1) | $Fe_3Sn$-like local environment | 0.39 | 0.05 | 24.4 | †0.30 | 5.2 | †3.57 |
| Fe(2) | Sn-containing Fe-rich bcc-like local environment | 0.18 | *0.00 | 29.4 | †0.30 | 2.7 | †3.57 |
| Fe(3) | bcc Fe(Sn)-like local environment | 0.15 | *0.00 | 33.0 | †0.30 | 1.5 | †3.57 |

($\chi^2$ = 1.71)

Amorphous $Fe_{0.78}Sn_{0.22}$, 33 nm

| Component | Assignment | δ (mm s$^{-1}$) | 2ε (mm s$^{-1}$) | $B_{hf}$ (T) | Γ (mm s$^{-1}$) | Δ$B$ (T) | ($x$) in 3:x:1:1:x:3 |
|---|---|---|---|---|---|---|---|
| Fe(4) | $FeSn_2$-like low-field local environment | 0.63 | -0.75 | 12.0 | †0.48 | 3.6 | †3.74 |
| Fe(1) | $Fe_3Sn$-like local environment | 0.24 | 0.06 | 25.4 | †0.48 | 2.4 | †3.74 |
| Fe(2) | Sn-containing Fe-rich bcc-like local environment | 0.14 | *0.00 | 29.4 | †0.48 | 2.3 | †3.74 |

($\chi^2$ = 1.08)

Crystalline $Fe_3Sn$, 25 nm

| Component | Assignment | δ (mm s$^{-1}$) | 2ε (mm s$^{-1}$) | $B_{hf}$ (T) | Γ (mm s$^{-1}$) | Δ$B$ (T) | ($x$) in 3:x:1:1:x:3 |
|---|---|---|---|---|---|---|---|
| Fe(5) | $Fe_5Sn_3$- or $Fe_3Sn_2$-like broad local environment | 0.25 | -0.24 | 19.5 | †0.28 | 8.2 | †3.70 |
| Fe(1) | $Fe_3Sn$-like local environment | 0.28 | 0.13 | 25.1 | †0.28 | 1.6 | †3.70 |
| Fe(3) | bcc Fe(Sn)-like local environment | 0.13 | *0.00 | 32.3 | †0.28 | 2.2 | †3.70 |

($\chi^2$ = 1.66)

**Table S1 | Best-fit Mössbauer parameters for amorphous $Fe_{0.78}Sn_{0.22}$ films and the crystalline $Fe_3Sn$ reference film at room temperature.**
Listed parameters include the isomer shift δ, quadrupole shift 2ε, hyperfine field $B_{hf}$, Lorentzian linewidth Γ, Gaussian width Δ$B$, sextet intensity ratio 3:x:1:1:x:3, area fraction, and goodness-of-fit $\chi^2$. Fixed parameters and common parameters are indicated as in the original fitting table.